\documentclass{SciPost}

\hypersetup{
    colorlinks,
    linkcolor={red!50!black},
    citecolor={blue!50!black},
    urlcolor={blue!80!black}
}

\usepackage[bitstream-charter]{mathdesign}
\usepackage{bm,bbm}
\usepackage[normalem]{ulem}
\usepackage{microtype}
\DeclareSymbolFont{usualmathcal}{OMS}{cmsy}{m}{n}
\DeclareSymbolFontAlphabet{\mathcal}{usualmathcal}

\fancypagestyle{SPstyle}{
\fancyhf{}

\fancyfoot[C]{\textbf{\thepage}}
}

\begin{document}

\pagestyle{SPstyle}

\begin{center}{\Large \textbf{\color{scipostdeepblue}{
Many-body effects beyond excitons in second-harmonic generation of monolayer MoS$_\mathbf{2}$\\
}}}\end{center}

\begin{center}\textbf{
Peio Garcia-Goiricelaya\textsuperscript{1,2$\star$} and
Julen Iba\~nez-Azpiroz\textsuperscript{3,4,5$\dagger$}
}\end{center}

\begin{center}
{\bf 1} Department of Physics, University of the Basque Country UPV/EHU,\\ 48940 Leioa, Basque Country, Spain
\\
{\bf 2} EHU Quantum Center, University of the Basque Country UPV/EHU,\\ 48940, Leioa, Bizkaia, Spain
\\
{\bf 3} Centro de F\'{i}sica de Materiales, Universidad del Pa\'{i}s Vasco UPV/EHU,\\ 20018 Donostia-San Sebasti\'{a}n, Spain
\\
{\bf 4} Ikerbasque Foundation, 48013 Bilbao, Spain
\\
{\bf 5} Donostia International Physics Center, 20018 Donostia-San Sebasti\'{a}n, Spain
\\[\baselineskip]
$\star$ \href{mailto:peio.garcia@ehu.eus}{\small peio.garcia@ehu.eus}\,,\quad
$\dagger$ \href{mailto:julen.ibanez@ehu.eus}{\small julen.ibanez@ehu.eus}
\end{center}

\section*{\color{scipostdeepblue}{Abstract}}
\textbf{\boldmath{%
We present a quantitative study of many-body effects including the three-particle level on second-harmonic generation in monolayer MoS$_2$. Our approach combines many-body perturbation theory with time-dependent current-density-functional theory within an \textit{ab initio} framework in the optical limit. The inclusion of two-particle excitonic effects \textit{via} a dynamical long-range linear exchange–correlation kernel reproduces the qualitative features of the second-harmonic response, but differs from the experimentally reported magnitude by nearly a factor of two. By incorporating three-particle trionic correlations through a static long-range quadratic exchange-correlation kernel, we achieve significantly improved quantitative agreement with experiment for the three main bound excitonic resonances. These findings suggest that many-body interactions beyond the two-body excitonic level can play an important role in accurately describing second-order optical responses in two-dimensional semiconductors.
}}

\vspace{\baselineskip}

\vspace{10pt}
\noindent\rule{\textwidth}{1pt}
\tableofcontents
\noindent\rule{\textwidth}{1pt}
\vspace{10pt}


\section{Introduction}\label{sec:intro}
Nonlinear optics encompasses physical phenomena in which the response 
of materials to incident light is not linearly proportional to the field amplitude~\cite{boyd2008nonlinear}.
Unlike the linear regime, nonlinear optical processes enable the generation of light at modified frequencies, for instance, through second-harmonic generation (SHG)~\cite{PhysRevLett.7.118}.
The last decade has witnessed a growing interest in higher-order optical responses for their technological potential.
A notable example is the bulk photovoltaic effect (BPVE) that occurs in noncentrosymmetric crystals~\cite{Sturman1992}, investigated as a promising mechanism for direct current generation beyond conventional solar cells~\cite{Spanier2016}.
Similarly, nonlinear optical effects in topological materials have gained significant interest~\cite{Wu2017,PhysRevB.98.165113,Osterhoudt2019,Ma2019,Ni2020} due to their deep connection with topological Berry-phase-like quantities~\cite{doi:10.1126/sciadv.1501524}.

A robust and predictive theoretical understanding of nonlinear optics is therefore highly desirable for the development of emerging optical technologies~\cite{Luppi_2016}.
In this context, first principles calculations based on density functional theory (DFT) play a central role, providing  microscopic and system-specific insights~\cite{PhysRevLett.109.116601,PhysRevLett.121.176604,PhysRevB.100.064301,PhysRevB.100.195305,PhysRevB.107.205204,h8rp-rtn8}.
However, achieving quantitative accuracy requires the explicit incorporation of many-body interactions into these theoretical frameworks~\cite{martin2016interacting}.
This is particularly important in low-dimensional semiconducting systems, where reduced electronic screening amplifies many-particle correlated effects~\cite{DESLIPPE2016}.
The emergence of collective excitations often dominates the optical response of such nanostructured materials, which have rapidly become leading platforms for future technological applications~\cite{https://doi.org/10.1002/adma.201605886}.

A clear view of the role of many-body interactions in nonlinear optical responses can be obtained within the framework of time-dependent DFT (TDDFT)~\cite{PhysRevLett.52.997}.
In this formalism, the interacting quadratic density response function is expressed through a Dyson-like shorthand equation as~\cite{Gross1996}
\begin{equation}\label{tddft2}
 \chi_{\rho2}=\varepsilon^{-1}\chi^{0}_{\rho2}\varepsilon^{-1}\varepsilon^{-1}+\chi_{\rho1}g_{\mathrm{xc}}\chi_{\rho1}\chi_{\rho1},
\end{equation}
where the linear contribution is
\begin{equation}\label{tddft1}
 \chi_{\rho1}=\chi^{0}_{\rho1}\varepsilon^{-1},
\end{equation}
with $\varepsilon^{-1}=1+(v_{\mathrm{c}}+f_{\mathrm{xc}})\chi_{\rho1}$ the inverse of the dielectric function and $v_{\mathrm{c}}$ the Coulomb interaction, \textit{i.e.}~the functional derivative of the Hartree potential with respect to density.
Above, $\chi_{\rho n}$ and $\chi^{0}_{\rho n}$ denote the 
many-body (interacting) and independent-particle (non-interacting) density response functions at $n$-th order, respectively.
In turn, $f_{\mathrm{xc}}$ and $g_{\mathrm{xc}}$ are the \textsl{unknown} linear and quadratic exchange-correlation (xc) kernels, \textit{i.e.}~the first and second functional derivatives of the xc potential with respect to density, respectively.
As seen by the equations above, these terms drive the many-body renormalization of linear and quadratic optical responses and are therefore the primary focus of this study.
From a many-body perturbation theory (MBPT) perspective, Eqs.~(\ref{tddft1}) and (\ref{tddft2}) are formally equivalent to contracting the correlation parts of the two- and three-particle Green’s functions, respectively~\cite{PhysRevA.83.062122}.
Importantly, this establishes a correspondence between the two- and three-body interactions and the linear $f_{\mathrm{xc}}$ and quadratic $g_{\mathrm{xc}}$ xc kernels of TDDFT, respectively.
\begin{figure}
  \includegraphics[width=\linewidth]{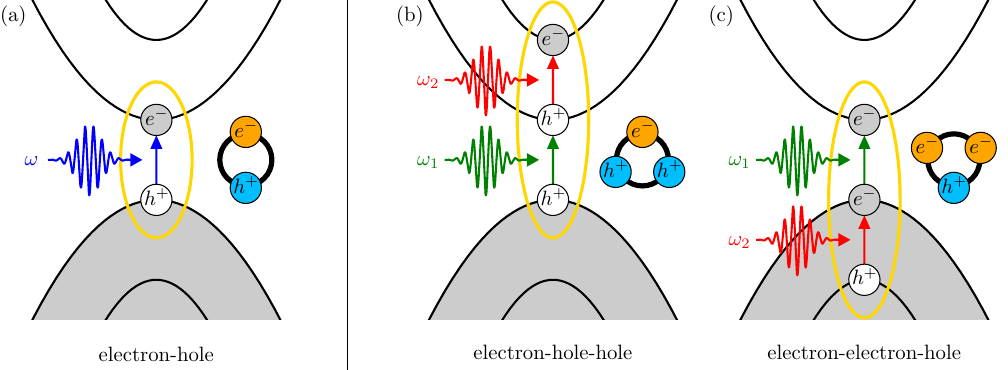}
  \caption{Schematic of two-particle (a) and three-particle (b,c) many-body interactions in the linear and quadratic optical response of a semiconductor, respectively. Grey and white shaded lines represent valence and conduction bands, respectively.}
  \label{fig:fig1}
\end{figure}

When evaluating the linear response to light, the absorption of a photon with frequency $\omega$ excites an electron from the valence band to the conduction band, leaving behind a hole, as schematically illustrated in Fig.~\ref{fig:fig1}(a).
Within TDDFT, $f_{\mathrm{xc}}$ captures the resulting two-body electron-hole interaction, which can lead to the formation of a two-particle bound state, known as an \emph{exciton}~\cite{PhysRev.37.17,gross_excitons}.
At second order, the consecutive absorption of two photons with frequencies $\omega_{1}$ and $\omega_{2}$ excites either one electron and two holes or two electrons and one hole, as depicted in Figs.~\ref{fig:fig1}(b) and (c), respectively.
In this case, $g_{\mathrm{xc}}$ incorporates the resulting three-body electron-electron-hole and electron-hole-hole interactions, which can lead to the formation of a three-particle bound states, known as \emph{trions} or charged excitons~\cite{PhysRevLett.1.450,keldysh_trions}.
Such composite quasiparticles have been experimentally observed in photoemission spectra emerging from neutral excitons upong carrier doping~\cite{PhysRevLett.71.1752}.
However, even in undoped semiconductors that do not host bound trions, three-particle trionic correlations can still renormalize quadratic optical responses through the second-order xc kernel $g_{\mathrm{xc}}$, as evident from Eq.~(\ref{tddft2}).

The most direct way of computing nonlinear optical responses beyond the independent-particle approximation is to evaluate the response functions explicitly in terms of many-body wave functions and energies.
Over the past two decades, such calculations have been extensively carried out at second order using a basis of two-particle excitonic states~\cite{PhysRevB.65.035205,PhysRevB.71.195209,PhysRevB.89.081102,PhysRevB.90.199901,PhysRevB.89.235410,PhysRevB.92.235432,PhysRevB.96.235206,PhysRevB.97.205432,PhysRevB.101.045104,PhysRevB.103.075402,PhysRevB.104.235203,10.1073/pnas.1906938118,PhysRevB.108.075413,doi:10.1021/acs.nanolett.4c03434,Esteve-Paredes2025}, while systematically neglecting three-particle trionic contributions.
In parallel, state-of-the-art TDDFT approaches typically start from a quasiparticle-corrected independent-particle response and incorporate excitonic effects through approximate long-range forms of the linear xc kernel, whereas trionic effects encased in the quadrtatic xc kernel are always omitted~\cite{https://doi.org/10.1002/pssb.200983954,10.1063/1.3457671,PhysRevB.82.235201,PhysRevB.83.115205,10.1063/1.5126501,PhysRevB.107.205101}.
As evident from Eq.~(\ref{tddft2}), this leads to a partially interacting second-order response that accounts only for two-particle correlations, again excluding three-particle contributions.
This constitutes a significant limitation: just as the two-body electron-hole interaction is essential for describing linear optical absorption, nonlinear optical responses are likewise expected to be increasingly governed by many-body interactions beyond the two-particle level.

In this work, we investigate the influence of many-body interactions including the three-particle level on SHG.
We have selected monolayer MoS$_{2}$ as a case study for two main reasons.
First, its reduced dimensionality gives rise to pronounced many-body excitonic~\cite{PhysRevB.86.115409,PhysRevB.88.045412,PhysRevLett.111.216805,PhysRevB.93.235435,PhysRevB.94.155406} and trionic~\cite{Ross2013,Mak2013} signatures in optical absorption and photoluminescence spectra.
Second, this material, as well as similar transition-metal dichalcogenide compounds, have recently shown signatures of remarkable nonlinear optical processes~\cite{PhysRevB.87.201401,PhysRevB.90.121409,PhysRevB.92.159901,zhang_enhanced_2019,PhysRevB.108.165418}. 
Our main result shows that including quadratic xc contributions that account for trionic effects leads to an improved quantitative agreement with experimental SHG spectra, emphasizing the important role of three-body interactions in accurately modeling second-order photoresponses.

The paper is organized as follows.
In Sec.~\ref{sec:theory}, we present the main theoretical framework.
We first introduce the second-harmonic generation tensor renormalized by many-body interactions.
We then describe in detail the modeling of monolayer MoS$_{2}$.
In Secs.~\ref{sec:linear} and~\ref{sec:shg}, we present and analyze the numerical results for the linear and quadratic optical regimes, respectively.
We conclude in Sec.~\ref{sec:summary} with a summary and discussion.
In Appendices~\ref{sec:comp} and \ref{sec:apptheory} we provide the computational details of the calculations and describe our theoretical methodology, respectively.
Appendix~\ref{sec:apptrions} is devoted to explaining how trionic peaks can be described within our theoretical framework.
\section{Theoretical framework}
\label{sec:theory}
\subsection{Tensorial many-body description of second-harmonic generation}
\label{subsec:}
Eqs.~(\ref{tddft1}) and~(\ref{tddft2}) describe the interacting linear and quadratic responses within TDDFT, respectively.
Although formally appealing, their scalar formulation makes it cumbersome to resolve directional components of the response, which are more naturally captured within a tensorial framework.
To this end, we employ time-dependent current-density functional theory (TDCDFT)~\cite{PhysRevLett.77.2037,PhysRevLett.79.4878}.
This formalism represents the tensorial extension of TDDFT, in which the charge density and scalar potentials are replaced by the current density and electric fields, and the scalar response functions and xc kernels by their tensorial counterparts.
In addition, TDCDFT allows the response equations to be formulated without long-wavelength divergences in the optical limit and without divergences at zero frequency in the static limit.
Details of the connection between the two frameworks are provided in Appendix~\ref{sec:apptheory}.

In this work, we focus on SHG, \textit{i.e.~}the quadratic optical process in which two photons of frequency $\omega$ are absorbed to generate radiation at frequency $2\omega$, corresponding to $\omega_{1}=\omega_{2}=\omega$ in Figs.~\ref{fig:fig1}(b) and (c).
In this case, the interacting quadratic optical response at the microscopic level can be expressed in terms of polarizability tensors as~\cite{PhysRevB.107.205101}
\begin{equation}\label{eq:alphashg}
 \begin{split}
  \bm{\alpha}_{2}(\omega,\omega)=\bm{\varepsilon}^{-1,T}(2\omega)\bm{\alpha}^{0}_{2}(\omega,\omega)\bm{\varepsilon}^{-1}(\omega)\bm{\varepsilon}^{-1}(\omega)+i\bm{\alpha}_{1}^{T}(2\omega)\bm{B}_{\mathrm{LRC}}(\omega,\omega)\bm{\alpha}_{1}(\omega)\bm{\alpha}_{1}(\omega)
 \end{split},
\end{equation}
where the linear contribution is given by
\begin{equation}\label{eq:alpha1}
\bm{\alpha}_{1}(\omega)=\bm{\alpha}^{0}_{1}(\omega)\bm{\varepsilon}^{-1}(\omega),
\end{equation}
with
\begin{equation}\label{eq:epsilon}
\bm{\varepsilon}^{-1}(\omega)=\bm{\mathbb{1}}-[4\pi-\bm{A}_{\mathrm{LRC}}(\omega)]\bm{\alpha}_{1}(\omega),
\end{equation}
which defines the inverse of the optical dielectric tensor.
Above, $\bm{\alpha_{n}}$ and $\bm{\alpha_{n}^{0}}$ are the many-body and independent-particle optical polarizability tensors at $n$-th order, respectively.
In Eq.~(\ref{eq:alphashg}), two-body excitonic effects enter through the optical dielectric tensor \textit{via} $\bm{A}_{\mathrm{LRC}}$, while three-body trionic effects are encoded in $\bm{B}_{\mathrm{LRC}}$.
These quantities are material-dependent tensors that parameterize the strength of the long-range contributions (LRC) to the linear and quadratic xc kernels in the optical limit (see Appendix~\ref{subsec:tdcdftopt}).
Due to the space group $P\overline{6}m2$ and the point group $D_{3h}$ of monolayer MoS$_2$, the in-plane symmetry-allowed components of second- and third-rank tensors satisfy the equivalence relations $xx = yy$ and $xxy=xyx=\allowbreak yxx=-yyy$, respectively.

Expressions similar to Eqs.~\ref{eq:alphashg} and \ref{eq:alpha1} were first derived in Ref.~\cite{PhysRevB.54.8540} in the context of nonlinear optical responses beyond the local-density approximation within a polarization-density-based formalism in the static limit.
Using Miller's conjecture~\cite{10.1063/1.1754022}, the contribution of three-particle correlations (see the second term on the right-hand side of Eq.~(\ref{eq:alphashg}) was estimated for several typical semiconductors to be on the order of 15\%--20\% of the total signal, suggesting that it is comparable to the contribution from local-field effects.

Our calculation strategy consists of several steps.
Based on DFT calculations, we first include one-body electron–electron and two-body electron–hole interactions within MBPT using the $GW$ approximation~\cite{PhysRev.139.A796,HEDIN19701} and the Bethe–Salpeter equation (BSE)~\cite{PhysRev.84.1232}, respectively.
This yields an accurate macroscopic optical dielectric tensor that incorporates quasiparticle corrections and excitonic effects.
From it, we extract the linear LRC screening tensor $\bm{A}_{\mathrm{LRC}}$, which reproduces the microscopic interacting linear optical response [see Eqs.~(\ref{eq:alpha1}) and (\ref{eq:epsilon})].
This tensor is then employed in state-of-the-art calculations of the microscopic interacting SHG response using Eq.~(\ref{eq:alphashg}), where the quantitative impact of trionic effects is assessed by tuning the quadratic LRC screening tensor $\bm{B}_{\mathrm{LRC}}$.
The connection between these microscopic quantities and their macroscopic counterparts is described in Appendix~\ref{subsec:macromicro}.
Technical details of the calculations can be found in Appendix~\ref{sec:comp}.
\subsection{Modeling monolayer MoS$_{\mathbf{2}}$}
\begin{figure}[t]
  \includegraphics[width=\linewidth]{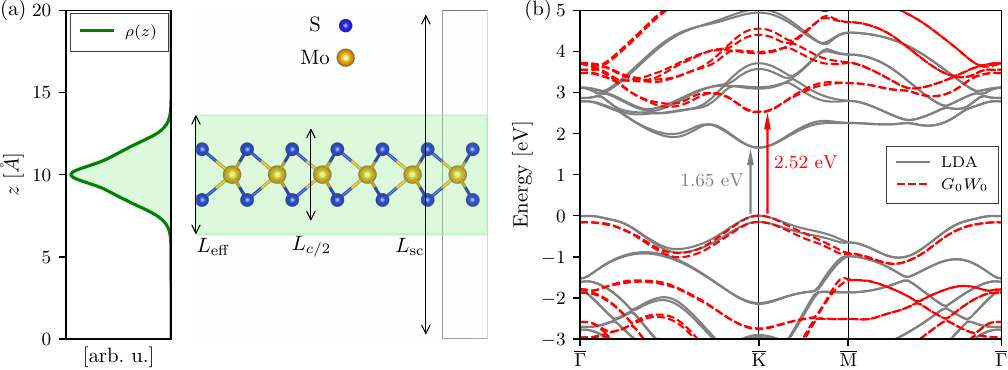}
  \caption{(a) Out-of-plane extension of the in-plane averaged charge density (left) together with the side view of the slab crystal unit (right) of monolayer MoS$_{2}$. Supercell thickness, effective thickness, and half of the bulk thickness are $L_{\mathrm{sc}}=20~\text{\AA}$, $L_{\mathrm{eff}}=7.3~\text{\AA}$ and $L_{c/2}=6.15~\text{\AA}$~\cite{young}, respectively. The green-shaded area highlights the effective out-of-plane spatial extent of the electrons. (b) LDA (solid grey lines) and $G_{0}W_{0}$ (dashed red lines) electronic band structure of monolayer MoS$_{2}$. LDA and $G_{0}W_{0}$ direct energy band gaps are equal to $1.65~\mathrm{eV}$ and $2.52~\mathrm{eV}$, respectively.}
  \label{fig:fig2}
\end{figure}
To model  monolayer MoS$_{2}$, we considered a supercell (sc) with a hexagonal close-packed structure of two planes of sulfur atoms with an intercalated plane of molybdenum atoms.
The in-plane lattice parameter is set equal to the experimental bulk value, \textit{i.e.}, $3.16~\text{\AA}$~\cite{young}, and a supercell height of $L_{\mathrm{sc}}=20~\text{\AA}$ is used in the out-of-plane direction, providing a sufficiently large vacuum region to avoid spurious interactions between periodic images.
Spurious Coulomb interactions are further suppressed using the truncated scheme of Ref.~\cite{PhysRevB.96.075448}.
The equilibrium atomic positions were determined by relaxing all atomic forces up to at least $10^{-6}~\text{Ry/}a_{0}$, leading to a relaxed interplanar distance equal to $1.55~\text{\AA}$, equivalent to a Mo-S bond length of $2.39~\text{\AA}$, in good agreement with previous theoretical calculations~\cite{PhysRevB.84.155413,PhysRevB.101.054304}.

Following a well-established procedure~\cite{PhysRevB.97.245143}, the response tensors are rescaled from the supercell to the monolayer by the factor $L_{\mathrm{sc}}/L_{\mathrm{eff}}$, where $L_{\mathrm{eff}}$ is the effective monolayer thickness, yielding 3D-like quantities.
We used a value of $L_{\mathrm{eff}}=7.3~\text{\AA}$, obtained from the spatial extent of the charge density along the out-of-plane direction, as shown in Fig.~\ref{fig:fig2}(a).
This value is in close agreement with the experimental atomic thickness of approximately $8~\text{\AA}$ reported in the Supplemental Information of Ref.~\cite{PhysRevB.87.201401}.
We note that the effective thickness employed here differs from the value $L_{c/2}=6.15~\text{\AA}$ used in other works, which corresponds to half of the bulk out-of-plane lattice parameter~\cite{young} [see Fig.~\ref{fig:fig2}(a)].
\section{Electronic band structure and linear optical response}
\label{sec:linear}
We begin by inspecting the electronic band structure of monolayer MoS$_{2}$.
The LDA calculation shown by solid grey lines in Fig.~\ref{fig:fig2}(b) predicts that the system is a semiconductor with a direct energy band gap of about $1.65~\mathrm{eV}$ at the high-symmetry point $\overline{\mathrm{K}}$.
In order to improve this description, we include quasiparticle corrections to the band structure by approximating self-energy effects of single-particle excitations arising from electron–electron interactions within the one-shot $G_{0}W_{0}$ approximation.
The resulting quasiparticle band dispersion is illustrated by the dashed red lines in Fig.~\ref{fig:fig2}(b), which predicts a direct band gap of $2.52~\mathrm{eV}$, much closer to the experimentally reported value of $\sim2.34~\mathrm{eV}$~\cite{doi:10.1021/acsnano.9b03652}.

\begin{figure}[t]
  \includegraphics[width=\linewidth]{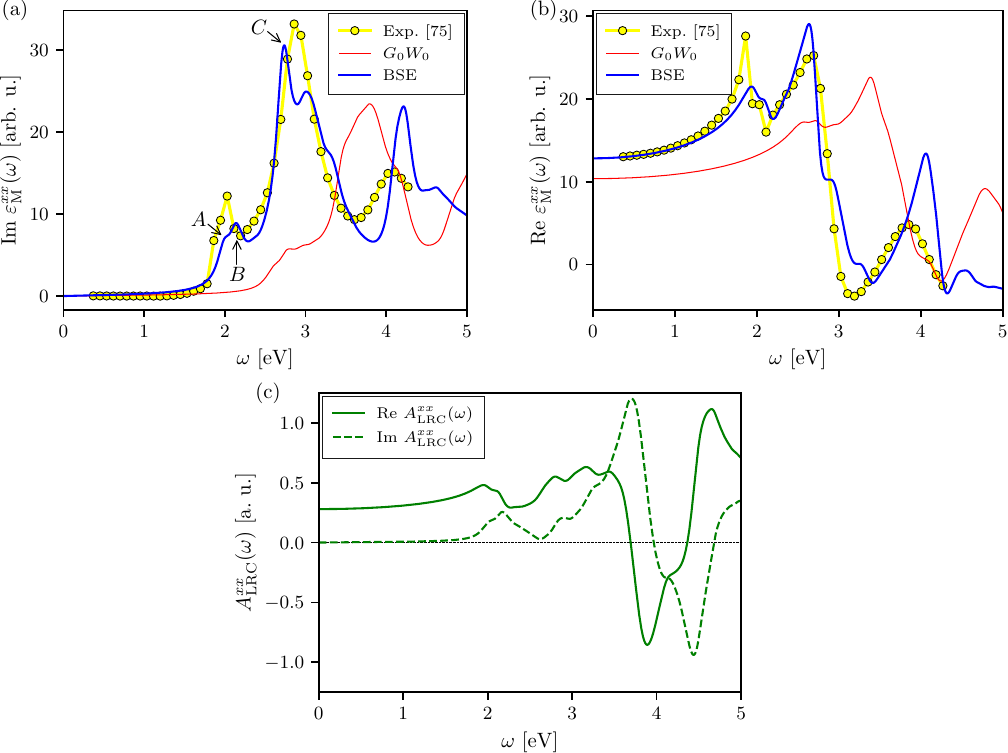}
  \caption{(a) Imaginary and (b) real parts of the macroscopic optical dielectric spectrum of monolayer MoS$_{2}$. Yellow lines with circles represent experimental data from Ref.~\cite{Ermolaev2020}. Narrow red and broad blue lines represent our calculations within the $G_{0}W_{0}$ and BSE pictures, respectively. (c) In-plane component of the linear LRC screeening tensor $\bm{A}_{\mathrm{LRC}}(\omega)$. Solid and dashed green lines represent the real and imaginary parts, respectively.}
  \label{fig:fig3}
\end{figure}
We next focus on the linear optical response.
In order to capture excitonic effects of two-particle excitations arising from electron-hole interactions, we solve BSE to obtain the macroscopic optical dielectric spectrum.
In Figs.~\ref{fig:fig3}(a) and (b), we compare the imaginary and real parts of this quantity, respectively, as calculated within the $G_{0}W_{0}$ and BSE approaches, shown by the narrow red and broad blue lines, respectively.
The BSE results exhibit very good agreement with the experimental data~\cite{Ermolaev2020}, shown by yellow lines with circles, and clearly improve upon the $G_{0}W_{0}$ spectrum by incorporating two-body excitonic effects.
Indeed, the BSE optical spectrum captures the spin-split $A$ and $B$ exciton double-peak structure between $2$ and $2.15~\mathrm{eV}$, followed by the $C$ peak near $2.7~\mathrm{eV}$.
All these results are in good agreement with previous DFT plus MBPT calculations~\cite{PhysRevB.86.115409,PhysRevB.88.045412,PhysRevLett.111.216805,PhysRevB.93.235435,PhysRevB.94.155406}.

Based on the BSE results shown in Figs.~\ref{fig:fig3}(a) and (b), we now proceed to construct a tensorial xc kernel in the linear regime that accounts for two-body excitonic effects within TDCDFT.
In the optical limit, the linear xc kernel of TDDFT is dominated by its long-range contribution and behaves \textsl{exactly} like an attractive screened Coulomb interaction that emulates the electron-hole attraction, in such a way that $\lim_{\mathbf{q}\to0}f_{\mathrm{xc}}(\mathbf{q})=-A_{\mathrm{LRC}}/q^2$, where $\mathbf{q}$ is the crystal momentum transfer and the screening parameter $A_{\mathrm{LRC}}$ is an \textsl{unknown} material-dependent quantity~\cite{PhysRevLett.88.066404,RevModPhys.74.601}.
Empirical~\cite{PhysRevB.68.205112,PhysRevB.69.155112,PhysRevB.95.205136} and nonempirical~\cite{PhysRevLett.107.186401,PhysRevLett.107.216402,PhysRevB.87.205143,PhysRevLett.114.146402,PhysRevLett.115.137402,PhysRevLett.117.159701,PhysRevLett.117.159702} static evaluations of the linear LRC screening parameter have only proven to work well to some extent for \textit{continuum} excitonic effects and at most one bound exciton peak, but fail to describe multiple excitonic features.
On the other hand, a dynamical extension of this quantity has been shown to be essential in order to reproduce subtle excitonic features within TDDFT starting from very good evaluations of the macroscopic optical dielectric function that can be, for instance, the result of BSE~\cite{PhysRevB.67.045207,PhysRevB.72.125203}.

As for TDDFT, an expression can be derived for calculating the linear LRC screening tensor within TDCDFT in the optical limit, in such a way that
\begin{equation}\label{eq:alrc}
 \bm{A}_{\mathrm{LRC}}(\omega)=-4\pi\left\{\left[\bm{\varepsilon}_{\mathrm{BSE}}(\omega)-\bm{\mathbb{1}}\right]^{-1}-\left[\bm{\varepsilon}_{\text{RPA}}(\omega)-\bm{\mathbb{1}}\right]^{-1}\right\},
\end{equation}
where $\bm{\varepsilon}_{\mathrm{RPA}}$ denotes the optical dielectric tensor in the random phase approximation (RPA), \textit{i.e.}~neglecting many-body interactions, and $\bm{\varepsilon}_{\mathrm{BSE}}$ the BSE optical dielectric tensor including excitonic effects.
It is worth noting that an equivalent expression to Eq.~(\ref{eq:alrc}) was first derived in Ref.~\cite{PhysRevB.54.8540} within a polarization-density based formalism in the static limit and was subsequently adopted in Refs.~\cite{PhysRevB.67.045207,PhysRevB.72.125203} within TDDFT.
The in-plane component of the linear LRC screening tensor calculated in this way is shown in Fig.~\ref{fig:fig3}(c).
The figure shows a rather intricate structure, with multiple peaks and dips in both the real and imaginary parts of $A^{xx}_{\mathrm{LRC}}(\omega)$ in the excitonic frequency region.
Importantly, this fine structure going beyond what can be achieved by simple approaches allows to incorporate the multipeak structure of excitons into TDCDFT.
\section{Two- and three-body effects on second-harmonic generation}\label{sec:shg}
\begin{figure}[t]
  \centering\includegraphics[width=\linewidth]{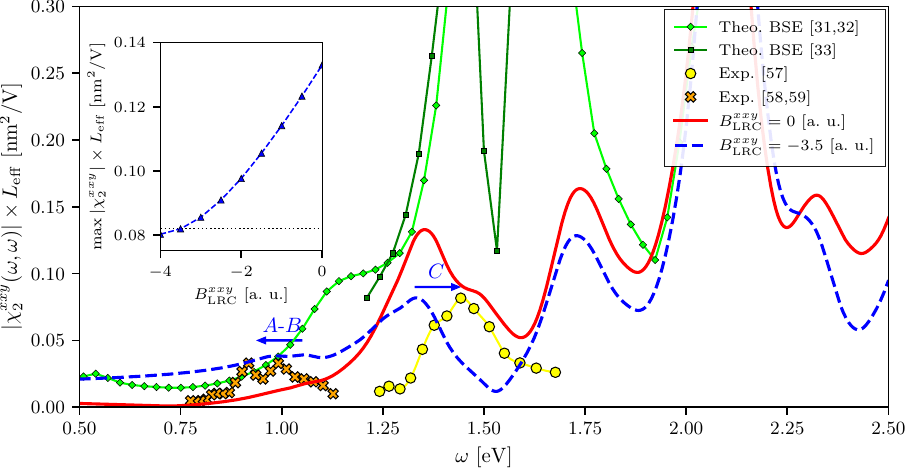}
  \caption{Absolute value of the macroscopic two-dimensional SHG susceptibility per unit area of monolayer MoS$_{2}$. The yellow line with circles and the orange line with crosses represent the experimental data from Refs.~\cite{PhysRevB.87.201401} and \cite{PhysRevB.90.121409,PhysRevB.92.159901}, respectively. The light and dark green lines with diamonds and squares represent the theoretical results including excitonic effects at the BSE level from Refs.~\cite{PhysRevB.89.081102,PhysRevB.90.199901} and \cite{PhysRevB.89.235410}, respectively. The solid red line represents our theoretical results including many-body interactions up to the two-particle level, \textit{i.e.}~excitonic effects with $B^{xxy}_{\mathrm{LRC}}=0~\mathrm{a.u.}$, while the dashed blue line includes also at the three-particle level, \textit{i.e.}~trionic effects with $B^{xxy}_{\mathrm{LRC}}=-3.5~\mathrm{a.u.}$. The inset shows the theoretical peak height at half the $C$-exciton energy, close to $1.5~\mathrm{eV}$, as a function of the in-plane longitudinal component of the quadratic LRC screening tensor $\bm{B}_{\mathrm{LRC}}$, with the dotted line indicating the experimental value from Ref.~\cite{PhysRevB.87.201401}.}
  \label{fig:fig4}
\end{figure}
We are now in a position to address our central objective of calculating the macroscopic SHG susceptibility from first principles, which is obtained as a macroscopic average of the interacting SHG response at the microscopic level (see Appendix~\ref{subsec:macromicro}).
We evaluate the response by gradually turning on many-body interactions.
In Fig.~\ref{fig:fig4}, we compare the calculated absolute value of the macroscopic SHG susceptibility obtained by including only excitonic effects at the two-particle level with $\bm{B}_{\mathrm{LRC}}=0~\mathrm{a.u.}$~(solid red line) with theoretical BSE results reported in the literature~\cite{PhysRevB.89.081102,PhysRevB.89.235410,PhysRevB.90.199901} (light and dark green lines with squares and diamonds, respectively), as well as with available experimental data from Refs.~\cite{PhysRevB.90.121409,PhysRevB.92.159901} near half the $A$- and $B$-exciton energies (orange line with crosses) and Ref.~\cite{PhysRevB.87.201401} near half the $C$-exciton energy (yellow line with circles).
The onset of the SHG signal occurs around $1~\mathrm{eV}$, coinciding with half the eigenvalues of the spin-split $A$ and $B$ bound excitons [see Fig.~\ref{fig:fig3}(a)].
The signal then reaches its first pronounced maximum through the strongly absorbing bound exciton $C$ [see Fig.~\ref{fig:fig3}(a)], as half of its eigenvalue coincides with the energy of the corresponding SHG peak (see Fig.~\ref{fig:fig4}).
Accordingly, our calculations predict a well-defined SHG resonance at $1.35~\mathrm{eV}$, which is in very good agreement with the experiment, showing a SHG peak slightly below $1.5~\mathrm{eV}$.
Although previous theoretical works have captured this overall trend~\cite{PhysRevB.89.081102,PhysRevB.89.235410,PhysRevB.90.199901}, our results additionally exhibit a remarkable quantitative agreement on the order of magnitude.
In contrast, earlier studies required an empirical rescaling of their spectra to reproduce the experimental magnitude of the SHG response.
This improvement arises from two main factors.
First, Wannier interpolation allows for Brillouin zone sampling to be several orders of magnitude denser than those employed in previous works, significantly enhancing numerical convergence and accuracy.
Second, we have adopted an effective layer thickness determined from the spatial extent of the charge density along the out-of-plane direction of the supercell, rather than the conventional choice of one half of the bulk lattice constant.
This definition is consistent with the experimentally measured atomic thickness~\cite{PhysRevB.87.201401} and provides a more physically grounded estimate.

Despite the substantial improvement over previous theoretical predictions, our SHG result for $\bm{B}_{\mathrm{LRC}}=0~\mathrm{a.u.}$~still yields an intensity of approximately $0.01$-$0.02~\mathrm{nm^{2}/V}$ near half the $A$- and $B$-exciton energies and $0.14~\mathrm{nm^{2}/V}$ near half the $C$-exciton energy, significantly below and above the corresponding experimental values of $0.04~\mathrm{nm^{2}/V}$ and $0.08~\mathrm{nm^{2}/V}$, respectively.
This difference in intensity between our calculations, which include only excitonic effects, and the experimental measurements can be attributed to several factors.
One possible source of the discrepancy is the substrate, which can modify the local electric field and induce interference effects not included in our formalism, thereby significantly altering the measured SHG intensity~\cite{Cheng_2019,Song:23}.
Nevertheless, the independent experimental studies reported in Refs.~\cite{PhysRevB.90.121409,PhysRevB.92.159901,PhysRevB.87.201401} calibrate the absolute value of the SHG susceptibility of the monolayer by removing substrate-induced effects according to the established procedure described therein.
Therefore, a direct comparison with our calculations for an isolated monolayer should be justified, since substrate effects have, in principle, been removed from the reported data.

Another possible source of  discrepancy is the treatment of thermal effects on the photoexcited carrier lifetimes.
In our calculations, these effects are incorporated phenomenologically through a constant broadening parameter within the relaxation-time approximation, as commonly adopted in state-of-the-art theoretical studies.
In reality, however, carrier lifetimes depend on the excitation energy, the relevant scattering mechanisms and the temperature.
Consequently, if the assumed lifetime is shorter (longer) than the actual one, the calculated resonant features become broader and less intense (narrower and more intense), while their spectral positions remain essentially unchanged.
One would therefore expect this effect to be qualitatively similar for all three excitonic resonances.
Instead, the calculated intensities of the $A$ and $B$ excitons are lower than the experimental values, whereas that of the $C$ exciton is higher, indicating that lifetime effects alone cannot account for the observed discrepancies.

Finally, and most relevant to the present work, another possible source of discrepancy is the neglect of many-body interactions at the three-particle level in our SHG calculations.
At this stage, our treatment has accounted for many-body interactions only up to the two-particle level, meaning that only the first term on the right-hand side (r.h.s.)~of Eq.~(\ref{eq:alphashg}) has been considered in the evaluation of the interacting microscopic SHG response.
However, it is the second term that accounts for many-body effects at the three-particle level, which are themselves linearly screened, as the quadratic LRC screening tensor $\bm{B}_{\mathrm{LRC}}$ correlates three interacting linear optical responses.
Therefore, it is reasonable to expect that if an effect is large at first order, it will also manifest prominently at second order.
This idea contradicts the common approximation in the literature of disregarding the quadratic xc kernel~\cite{https://doi.org/10.1002/pssb.200983954,10.1063/1.3457671,PhysRevB.82.235201,PhysRevB.83.115205,10.1063/1.5126501,PhysRevB.107.205101}.

In our formulation, we exploit the \textsl{exact} static behavior of the quadratic xc kernel of TDDFT in the optical limit, which has been previously derived from a second-order BSE for the three-particle correlation function by H\"ubener~\cite{PhysRevA.83.062122}.
In this limit, the long-range contribution dominates, such that $\lim_{\mathbf{q}\to0}g_{\mathrm{xc}}(\mathbf{q},\mathbf{q})=B_{\mathrm{LRC}}/q^{3}$~\cite{PhysRevA.83.062122}, where $B_{\mathrm{LRC}}$ is a material-dependent parameter.
Unlike conventional (first-order) BSE, its second-order counterpart has not yet been implemented in practical calculations; therefore, the tensor $\bm{B}_{\mathrm{LRC}}$ cannot be obtained in the same way as $\bm{A}_{\mathrm{LRC}}$.
As a result, no previous studies appear to have addressed its explicit evaluation.
Nevertheless, by comparing experimental spectra with theoretical results that include many-body effects only up to the two-particle level, a reasonable range of values for $\bm{B}_{\mathrm{LRC}}$ can be inferred, as we do next.
This approach is conceptually similar to early treatments of the linear xc kernel in the optical limit within linear-response TDDFT, where empirical values of the linear LRC screening parameter $A_{\mathrm{LRC}}$ were adjusted to reproduce experimentally observed excitonic peaks~\cite{PhysRevLett.88.066404,PhysRevB.69.155112}
(see also Ref.~\cite{Byun_2020} for a recent review).

We proceed by comparing the maximum SHG peak intensity near half the $C$-exction energy from our theoretical predictions with the experimental value~\cite{PhysRevB.87.201401} as a function of the in-plane component of $\bm{B}_{\mathrm{LRC}}$.
The results are shown in the inset of Fig.~\ref{fig:fig4}, indicating that for values $B^{xxy}_{\mathrm{LRC}}\in[-4,0]~\mathrm{a.u.}$, the impact on the maximum SHG response is significant, reducing it by nearly a factor of two.
In particular, our theoretical results match the experimental peak intensity
for $B^{xxy}_{\mathrm{LRC}}=-3.5~\mathrm{a.u.}$, from which we identify this value as optimal for the quadratic LRC screening parameter.
The corresponding SHG spectrum is shown in main Fig.~\ref{fig:fig4} (dashed magenta line).
Besides the peak intensity, the shape of the main SHG peak at $1.4~\mathrm{eV}$ is very well captured by our calculations, with only a $\sim0.1~\mathrm{eV}$ energy redshift of our computed result compared to the experiment, a discrepancy that lies within the typical uncertainty of $GW$ and BSE calculations within the MBPT framework.
More interestingly, the same value of $B^{xxy}_{\mathrm{LRC}}$, adjusted to reproduce the experimental peak near half the $C$-exciton energy~\cite{PhysRevB.87.201401}, also enhances the SHG signal near half the $A$- and $B$-exciton energies, improving the quantitative agreement in this energy range with an independent experimental study~\cite{PhysRevB.90.121409,PhysRevB.92.159901}.
Thus, three-body effects enhance the SHG signal near half the $A$- and $B$-exciton energies while suppressing it near half the $C$-exciton energy.
This behavior is consistent with the formation of bound trions arising from strong three-body interactions, as observed in gate-voltage-doped monolayer MoS$_{2}$ in the vicinity of the $A$- and $B$-exciton energies~\cite{Ross2013,Mak2013}, which are expected to contribute to the SHG response at half these energies.
The formation of such bound collective states naturally enhances the SHG signal in this energy region.
By contrast, no trion formation has been reported near the $C$-exciton energy in doped samples.
Consequently, when three-body trionic correlations are included, enhanced screening is expected to dominate in this spectral region, leading to a suppression of the SHG signal.

Our main result demonstrates that three-particle correlations can be comparable in magnitude to two-particle contributions and are therefore indispensable for quantitative theoretical predictions of nonlinear optical responses, such as SHG.
This should not come as a surprise, given that experimental reports consistently indicate that trionic effects play a significant role in these types of two-dimensional materials~\cite{Ross2013,Mak2013}.
\section{Conclusions and Outlook}\label{sec:summary}
In summary, our \textit{ab initio} results show that many-body interactions beyond the conventional two-particle excitonic level
can be important for a quantitative description of the SHG spectrum in monolayer MoS$_{2}$.

In the first level of approximation, we perform TDCDFT calculations using a dynamical long-range linear xc kernel derived from the BSE to capture two-body interactions. By employing highly dense $\mathbf{k}$-point meshes enabled by the Wannier interpolation technique, we accurately reproduce both the multiple-excitonic structure and the correct order of magnitude of the SHG response.
However, this is not sufficient to achieve quantitative agreement with experiment, as calculations including only two-body excitonic effects underestimate the intensity by nearly a factor of two near half the $A$- and $B$-exciton energies and overestimate it by a similar factor near half the $C$-exciton energy.
By incorporating three-body trionic effects through a static long-range quadratic xc kernel, our calculations achieve a much better agreement with the excitonic peak intensities reported in two independent SHG experiments by simultaneously fixing the incorrect trends of the two-body approximation.
These findings reveal that three-body interactions, which are almost universally neglected in standard theoretical treatments, can play an important role in second-order optical processes, comparable to other previously quantified effects such as substrate and 
thermal contributions~\cite{Cheng_2019,Song:23}.
More broadly, our work closes a quantitative gap between theory and experiment for SHG in low-dimensional semiconductors and establishes a framework for improved predictions of other quadratic optical responses in emerging materials.

Building on the present analysis of SHG, it is natural to ask what the role of three-body interactions is in other nonlinear optical responses.
Among these, the shift- and injection-current contributions to the BPVE~\cite{Sturman1992} are particularly compelling, given the significant attention this phenomenon has received in recent years.
Such quadratic optical processes occur when the absorbed two photons have opposite energies, \textit{i.e.~}$\omega_{1}=\omega=-\omega_{2}$ [see Figs.~\ref{fig:fig1}(b,c)].
Adapting Eq.~(\ref{eq:sigma2}) from Appendix~\ref{sec:apptheory} to these processes, the Dyson-like expression in the frequency domain that accounts for such interacting optical conductivity tensors at the microscopic level within TDCDFT reads as~\cite{PhysRevB.107.205101}
\begin{equation}\label{eq:sigmabpve}
 \begin{aligned}
  \bm{\sigma}_{2}(\omega,-\omega)={\bm{\varepsilon}^{-1}}^{T}(0)\bm{\sigma}^{0}_{2}(\omega,-\omega)\bm{\varepsilon}^{-1}(\omega)\bm{\varepsilon}^{-1}(-\omega)+i\bm{\sigma}_{1}^{T}(0)\frac{\bm{B}_{\mathrm{LRC}}(\omega,-\omega)}{\omega^{2}}\bm{\sigma}_{1}(\omega)\bm{\sigma}_{1}(-\omega)
 \end{aligned},
\end{equation}
where $\bm{\sigma}_{n}$ and $\bm{\sigma}^{0}_{n}$ are the many-body and independent-particle conductivity tensors at $n$-th order.
By inspecting Eq.~(\ref{eq:sigmabpve}), we note that in semiconductors of practical interest with a finite band gap, the first term on the r.h.s.~can be non zero, whereas the second term will always vanish.
This follows from the fact that, in undoped semiconductors, the lineal conductivity in the static limit is zero, that is, $\bm{\sigma}_{1}(0)=0$.
As a result, when analyzing the shift- and injection-current contributions to BPVE, many-body effects enter only by means of the first term, thus limited to the two-particle excitonic level.
In this way, previous calculations based on excitonic wave functions and energies can be regarded as formally consistent in their treatment of many-body interactions~\cite{PhysRevB.101.045104,PhysRevB.103.075402,PhysRevB.104.235203,10.1073/pnas.1906938118,PhysRevB.108.075413,Esteve-Paredes2025,doi:10.1021/acs.nanolett.4c03434}.
However, this result appears to be rather specific, owing to the direct current (DC) nature of BPVE, which manifests through the DC term $\bm{\sigma}_{1}(0)$ in Eq.~(\ref{eq:sigmabpve}).
In general, three-particle correlations to quadratic responses cannot be expected to vanish; therefore, theoretical frameworks such as the present one that explicitly include them are necessary.
\section*{Acknowledgements}
The authors are grateful to Alejandro Molina-S\'anchez for helpful discussions related with $GW$ \& BSE calculations, as well as guidance using the \textsc{Yambo} code.
The authors acknowledge financial support from the European Union's Horizon 2020 Research and Innovation Programme through the European Research Council (ERC) under Grant Agreement No.~946629, from the Spanish Ministry of Science, Innovation and Universities (MICIU) through the ``Proyectos de Generación de Conocimiento'' program under Grant No.~PID2023-147324NA-I00, and from the Department of Education, Universities and Research of the Basque Government through the ``Ikertalde 2026'' program under Grant No.~IT1837-26.
%
\begin{appendix}
\numberwithin{equation}{section}
\numberwithin{figure}{section}
\section{Computational details}
\label{sec:comp}
We performed ground-state electronic structure calculations using density functional theory (DFT)~\cite{dft1,dft2} in the local density approximation (LDA)~\cite{lda} as implemented in the \textsc{Quantum ESPRESSO} code package~\cite{qe1,qe2,qe3}.
The interaction between valence electrons and atomic cores was modeled by means of norm-conserving pseudopotentials~\cite{ncpp1,ncpp2} from the optimized norm-conserving Vanderbilt SG15 pseudopotentials database~\cite{oncv1,oncv2} including fully relativistic corrections~\cite{oncvso}.
In the case of molybdenum, besides $4d$ and $5s$ valence electrons, $4s$ and $4p$ semi-core electrons were also explicitly included as valence states, since they are known to be crucial for proper self-energy calculations based on the $GW$ method~\cite{PhysRevB.88.045412,MOLINASANCHEZ2015554}.
Self-consistent DFT calculations were carried out using a $\mathbf{k}$-point mesh of $12\times12$ in combination with fixed occupation values and a plane-wave basis set with a cutoff energy of $100~\text{Ry}$.

Starting from LDA wave functions and energies, we calculated quasiparticle (QP) corrections due to electron-electron interactions within the one-shot $G_{0}W_{0}$ approximation~\cite{PhysRev.139.A796,HEDIN19701} and, subsequently, linear optical absorption spectra including excitonic effects through BSE~\cite{PhysRev.84.1232}, as implemented in the \textsc{Yambo} code package~\cite{yambo1,yambo2}.
Concerning $G_{0}W_{0}$ calculations, we selected the main parameters by focusing on the convergence of the QP band-gap energy, as well as the conduction-band minimum and valence-band maximum, with a threshold  below $15~\text{meV}$.
We applied the plasmon-pole approximation~\cite{Lundqvist19671,Lundqvist19672,PhysRevB.3.1888} with an energy cutoff of $12~\mathrm{Ry}$ and $40~\mathrm{Ry}$ for the dielectric matrix and both the exchange and correlation parts of the self-energy, respectively, and we included up to 220 bands in the sum-over-states of the dielectric matrix and the correlation part of the self-energy.
We obtained converged values for a $\mathbf{k}$-point mesh of $30\times30$.
To speed up the convergence with respect to empty states, we adopted the technique described in Ref.~\cite{PhysRevB.78.085125}.
In addition, to accelerate convergences with respect to the $\mathbf{k}$-point sampling avoiding long-wavelength divergences of the electronic potential, we used the so-called random integration method of the screened potential (RIM-W)~\cite{Guandalini2023} together with the truncation of the Coulomb potential in the slab geometry.
In this way, we performed Monte Carlo integrations with $2\times10^6$ random points taking into account $\mathbf{G}$ vectors up to $2.5~\mathrm{Ry}$ and $1~\mathrm{Ry}$ for the bare and screened Coulomb potentials, respectively.
For the BSE calculations, we employed the Lanczos-Haydock solver~\cite{HAYDOCK1980215}, since our interest lies in the excitonic optical spectrum rather than in storing the full excitonic wave functions.
An energy cutoff of $25~\mathrm{Ry}$ and $8~\mathrm{Ry}$ is used for the electron-hole exchange and attraction parts, respectively, of the BSE kernel, and 14 occupied and 22 unoccupied states have been used to build up the excitonic Hamiltonian.
We used a $\mathbf{k}$-point mesh of $33\times33$, which is enough to provide a good convergence in the position of the excitonic peaks.

As a next step, we evaluated independent-particle SHG susceptibility tensors at LDA and $G_{0}W_{0}$ levels in the optical and long-wavelength limit.
We considered the length-gauge expressions~\cite{PhysRevB.48.11705,PhysRevB.52.14636,PhysRevB.61.5337,PhysRevB.97.245143} as implemented in the \textsc{Wannier90} code package~\cite{Pizzi2020,PhysRevB.107.205101}. 
To this end, starting from a set of 48 spin-split bands, we constructed 36 disentangled maximally localized Wannier functions  spanning the 14 high-energy valence bands and the 22 low-energy conduction bands using two $p$ trial orbitals centered on S atoms, as well as one $s$ and one $d$ trial orbitals centered on the Mo atom.
To obtain well-converged optical spectra, we used a dense $\mathbf{k}$-point interpolated mesh of $1000\times1000$.
With respect to the imaginary part of the complex $\hbar\tilde{\omega}$, we set $\eta=0.1~\mathrm{eV}$, consistent with photoexcited carrier lifetimes $(\sim10~\mathrm{fs})$~\cite{doi:10.1021/acs.nanolett.3c00732}.
Regarding the auxiliary parameter for regularizing energy denominators, we chose $\eta_{r}=0.05~\mathrm{eV}$.
The occupation factors and their derivatives were evaluated at zero temperature $(T=0~\mathrm{K})$.
\section{Connection between TDDFT and TDCDFT}\label{sec:apptheory}
In this appendix, we review the main derivation steps for obtaining tractable microscopic response equations within TDCDFT in the optical limit starting from TDDFT.
The connection between the macroscopic and microscopic levels of the response is also discussed.
\subsection{Time-dependent density functional theory}\label{subsec:tddft}
In TDDFT, one investigates the density response of a many-body system of electrons interacting \textit{via} the Coulomb interaction to an externally applied time-dependent potential.
Alternatively, one can adopt the auxiliary independent-particle system of non-interacting electrons, where the density response is now viewed with respect to the total time-dependent potential felt by the particles. 
For quadratic responses, this entails expanding the time-dependent density response of the interacting system in a power series of the external potential up to second order,
\begin{equation}
 \rho(1)=\int^{1}_{0}\chi_{\rho1}(1,2)V_{\mathrm{ext}}(2)d2+\iint^{1}_{0}\chi_{\rho2}(1,2,3)V_{\mathrm{ext}}(2)V_{\mathrm{ext}}(3)d2d3,
\end{equation}
or, in the non-interacting case, in a power series of the total potential,
\begin{equation}
 \rho(1)=\int^{1}_{0}\chi^{0}_{\rho1}(1,2)V_{\mathrm{tot}}(2)d2+\iint^{1}_{0}\chi^{0}_{\rho2}(1,2,3)V_{\mathrm{tot}}(2)V_{\mathrm{tot}}(3)d2d3,
\end{equation}
where $\chi_{\rho n}$ and $\chi^{0}_{\rho n}$ are the many-body and independent-particle density response functions and we adopted the notation $(\mathbf{r}_{n},t_{n})\equiv(n)$ with $n$ a positive integer and with the total potential related to the external potential as
\begin{equation}\label{vtot}
 V_{\mathrm{tot}}(\mathbf{r},t)=V_{\mathrm{ext}}(\mathbf{r},t)+V_{\mathrm{H}}(\mathbf{r},t)+V_{\mathrm{xc}}(\mathbf{r},t).
\end{equation}
The Hartree potential as a function of the density response is given by
\begin{equation}\label{vh}
 V_{\mathrm{H}}(1)=\int^{1}_{0}v_{\mathrm{c}}(1,2)\rho(2)d2,
\end{equation}
with the Coulomb interaction $v_{\mathrm{c}}(1,2)=\delta(t_{1}-t_{2})/|\mathbf{r}_{1}-\mathbf{r}_{2}|$, and the xc potential up to second order is written as
\begin{equation}\label{vxc}
 V_{\mathrm{xc}}(1)=\int^{1}_{0}f_{\mathrm{xc}}(1,2)\rho(2)d2+\iint^{1}_{0}g_{\mathrm{xc}}(1,2,3)\rho(2)\rho(3)d2d3.
\end{equation}
Applying now the chain rule in the definition of the many-body density response functions and taking into account the definition of the independent-particle density response function as well as the relation between the total and external potentials, one arrives to Eqs.~\ref{tddft1} and \ref{tddft2} at first and second order, respectively.
\subsection{Time-dependent current-density functional theory}\label{subsec:tdcdft}
TDCDFT follows the same basics as TDDFT, but the difference is that we are now interested in the microscopic response of the current-density to an externally applied time-dependent electric field, in the case of the many-body system, or to the total time-dependent electric field, in the case of the independent-particle system.
From the mathematical point of view, this change of paradigm means that we are no longer interested in scalar quantities, but in vectors and tensors.
In this way, up to second order, the time-dependent current-density response can be written as
\begin{equation}
 \mathbf{j}(1)=\int^{1}_{0}\bm{\sigma}_{1}(1,2)\mathbf{E}_{\mathrm{ext}}(2)d2+\iint^{1}_{0}\bm{\sigma}_{2}(1,2,3)\mathbf{E}_{\mathrm{ext}}(2)\mathbf{E}_{\mathrm{ext}}(3)d2d3,
\end{equation}
or alternatively as
\begin{equation}
 \mathbf{j}(1)=\int^{1}_{0}\bm{\sigma}^{0}_{1}(1,2)\mathbf{E}_{\mathrm{tot}}(2)d2+\iint^{1}_{0}\bm{\sigma}^{0}_{2}(1,2,3)\mathbf{E}_{\mathrm{tot}}(2)\mathbf{E}_{\mathrm{tot}}(3)d2d3,
\end{equation}
where $\bm{\sigma}_{n}$ and $\bm{\sigma}^{0}_{n}$ are the many-body and independent-particle current-density response tensors at $n$-th order, and the total electric field is related to the external electric field as
\begin{equation}\label{etot}
 \mathbf{E}_{\mathrm{tot}}(\mathbf{r},t)=\mathbf{E}_{\mathrm{ext}}(\mathbf{r},t)+\mathbf{E}_{\mathrm{H}}(\mathbf{r},t)+\mathbf{E}_{\mathrm{xc}}(\mathbf{r},t).
\end{equation}
The Hartree electric field as a function of the current-density response is given by
\begin{equation}\label{eh}
 \mathbf{E}_{\mathrm{H}}(1)=\int^{1}_{0}\bm{K}_{\mathrm{H}}(1,2)\mathbf{j}(2)d2,
\end{equation}
where $\bm{K}_{\mathrm{H}}$ is the tensorial Hartree kernel, which represents the tensor extension of the Coulomb interaction.
In turn, the xc electric field up to second order is written as
\begin{equation}\label{exc}
 \mathbf{E}_{\mathrm{xc}}(1)=\int^{1}_{0}\bm{F}_{\mathrm{xc}}(1,2)\mathbf{j}(2)d2+\iint^{1}_{0}\bm{G}_{\mathrm{xc}}(1,2,3)\mathbf{j}(2)\mathbf{j}(3)d2d3,
\end{equation}
with $\bm{F}_{\mathrm{xc}}$ and $\bm{G}_{\mathrm{xc}}$ the linear and quadratic tensorial xc kernels, respectively, which can also be viewed as the tensor extension of $f_{\mathrm{xc}}$ and $g_{\mathrm{xc}}$.

As for the density response, applying the chain rule in the definition of the many-body current-density response tensors and taking into account the definition of the independent-particle current-density response tensors, as well as the relation between the total and external electric fields, the Dyson-like linear and quadratic current-density response equations are derived,
\begin{equation}\label{tdcdft1}
 \bm{\sigma}_{1}(1,2)=\int\bm{\sigma}^{0}_{1}(1,3)\bm{\varepsilon}^{-1}(3,2)d2,
\end{equation}
with $\bm{\varepsilon}^{-1}(1,2)=\bm{\mathbb{1}}+\int\left[\bm{K}_{\mathrm{H}}(1,3)+\bm{F}_{\mathrm{xc}}(1,3)\right]\bm{\sigma}(3,2)d3$ the inverse of the dielectric tensor, and
\begin{equation}\label{tdcdft2}
 \begin{split}
  \bm{\sigma}_{2}(1,2,3)=\iiint [&\bm{\varepsilon}^{-1,T}(1,4)\bm{\sigma}^{0}_{2}(4,5,6)\bm{\varepsilon}^{-1}(5,2)\bm{\varepsilon}^{-1}(6,3)\\&+\bm{\sigma}_{1}^{T}(1,4)\bm{G}_{\mathrm{xc}}(4,5,6)\bm{\sigma}_{1}(5,2)\bm{\sigma}_{1}(6,3) ]d4d5d6
 \end{split},
\end{equation}
respectively.
As can be seen directly, Eqs.~\ref{tdcdft1} and \ref{tdcdft2} are nothing but the tensor extension of Eqs.~\ref{tddft1} and \ref{tddft2}.
\subsection{TDCDFT in the optical and long-wavelength limit}\label{subsec:tdcdftopt}
Since we are interested in optical responses, it is convenient to adapt TDCDFT equations to the optical and long-wavelength limit, which assumes that the microscopic response remains constant in the length-scale of the crystal's unit cell.

Regarding the Hartree contribution, the TDDFT Hartree potential is defined in Eq.~(\ref{vh}), while the TDCDFT Hartree electric field is defined by Eq.~(\ref{eh}).
With the aid of Maxwell's equation $\mathbf{E}(\mathbf{r},t)=-\boldsymbol{\nabla}V(\mathbf{r},t)$ and the continuity equation $\boldsymbol{\nabla}\cdot\mathbf{j}(\mathbf{r},t)=-\partial_{t}\rho(\mathbf{r},t)$, one can reformulate the TDCDFT Hartree electric field in the wave-vector and frequency space starting from the TDDFT Hartree potential and compare it with its original definition.
In this way, the tensorial Hartree kernel can be expressed as
\begin{equation}
 \bm{K}_{\mathrm{H}}(\mathbf{q}_{1},\mathbf{q}_{2},\omega)=\frac{4\pi}{i\omega}\delta_{\mathbf{q}_{1}\mathbf{q}_{2}}\frac{\mathbf{q}_{1}\odot\mathbf{q}_{2}}{|\mathbf{q}_{1}||\mathbf{q}_{2}|},
\end{equation}
where $\mathbf{q}_{n}$ represents the crystal momentum transfer of the $n$-th particle.
In the optical limit, \textit{i.e.~}$\lim\mathbf{q}_{n}\to0$, the tensorial Hartree kernel reads then
\begin{equation}
 \bm{K}_{\mathrm{H}}(\omega)=\frac{4\pi}{i\omega}\bm{\mathbb{1}}.
\end{equation}

\sloppy As for the xc terms, we assume that the nonlocal long-range contribution (LRC) dominates over all other terms in the optical limit.
In this way, excitonic and trionic effects in the optical limit within TDDFT are included by means of \textsl{exact} behaviors of the linear and quadratic xc kernels $\lim_{\mathbf{q}_{n}\to0}f_{\mathrm{xc}}(\mathbf{q}_{1},\mathbf{q}_{2},\omega)=-A_{\mathrm{LRC}}(\omega)/q^2$~\cite{PhysRevLett.88.066404} and $\lim_{\mathbf{q}_{n}}g_{\mathrm{xc}}(\mathbf{q}_{1},\mathbf{q}_{2},\mathbf{q}_{3},\omega_{1},\omega_{2})=B_{\mathrm{LRC}}(\omega_{1},\omega_{2})/q^{3}$~\cite{PhysRevA.83.062122}, respectively.
Both limits are derived from the first-order BSE two-particle and second-order BSE three-particle correlation functions within MBPT, respectively.
Following the same procedure used for the optical tensorial Hartree kernel, we also derive the expressions of the linear and quadratic tensorial xc kernels in the optical limit, which reads as
\begin{equation}
 \bm{F}_{\mathrm{xc}}(\omega)=-\frac{\bm{A}_{\mathrm{LRC}}(\omega)}{i\omega},
\end{equation}
and
\begin{equation}
 \bm{G}_{\mathrm{xc}}(\omega_{1},\omega_{2})=\frac{\bm{B}_{\mathrm{LRC}}(\omega_{1},\omega_{2})}{i\omega_{1}\omega_{2}},
\end{equation}
respectively.
Thus, taking into account these definitions, one can already write the Dyson-like linear and quadratic current-density response equations in the optical limit as
\begin{equation}\label{eq:sigma1}
 \bm{\sigma}_{1}(\omega)=\bm{\sigma}^{0}_{1}(\omega)\bm{\varepsilon}^{-1}(\omega),
\end{equation}
with $\bm{\varepsilon}^{-1}(\omega)=\bm{\mathbb{1}}+\frac{4\pi-\bm{A}_{\mathrm{LRC}}(\omega)}{i\omega}\bm{\sigma}_{1}(\omega)$ the inverse of the optical dielectric tensor, and
\begin{equation}\label{eq:sigma2}
 \begin{split}
  \bm{\sigma}_{2}(\omega_{1},\omega_{2})=&\bm{\varepsilon}^{-1,T}(\omega_{1}+\omega_{2})\bm{\sigma}^{0}_{2}(\omega_{1},\omega_{2})\bm{\varepsilon}^{-1}(\omega_{1})\bm{\varepsilon}^{-1}(\omega_{2})\\&+\bm{\sigma}_{1}^{T}(\omega_{1}+\omega_{2})\frac{\bm{B}_{\mathrm{LRC}}(\omega_{1},\omega_{2})}{i\omega_{1}\omega_{2}}\bm{\sigma}_{1}(\omega_{1})\bm{\sigma}_{1}(\omega_{2})
 \end{split},
\end{equation}
respectively.
Here, the advantage of using TDCDFT in the optical limit becomes evident, as one obtains current–density response equations that are independent of the wave vector, in contrast to the divergences that arise in density-based response functions within TDDFT in the optical limit.
Nevertheless, divergences related to the frequency are still present.
These are especially cumbersome for semiconductors, which present zero conductivity as the frequency goes to zero, leading to an indeterminate form of 0/0.

In this case, the response equations can be reformulated as a function of the polarization density instead of the current density.
The optical polarization-density response with respect to the many-body system and to the independent-particle system is described as
\begin{equation}\label{eq:peqaEext}
 \mathbf{p}(\omega=\omega_{1}+\omega_{2})=\bm{\alpha}_{1}(\omega)\mathbf{E}_{\mathrm{ext}}(\omega)+\bm{\alpha}_{2}(\omega_{1},\omega_{2})\mathbf{E}_{\mathrm{ext}}(\omega_{1})\mathbf{E}_{\mathrm{ext}}(\omega_{2}),
\end{equation}
and
\begin{equation}\label{eq:peqa0Etot}
 \mathbf{p}(\omega=\omega_{1}+\omega_{2})=\bm{\alpha}^{0}_{1}(\omega)\mathbf{E}_{\mathrm{tot}}(\omega)+\bm{\alpha}^{0}_{2}(\omega_{1},\omega_{2})\mathbf{E}_{\mathrm{tot}}(\omega_{1})\mathbf{E}_{\mathrm{tot}}(\omega_{2}),
\end{equation}
respectively, where $\bm{\alpha}_{n}$ and $\bm{\alpha}^{0}_{n}$ are the many-body and independent-particle polarizability tensors at $n$-th order, respectively.
In the absence of magnetization and free charge and current densities, the current and polarization-density vectors are related by $\mathbf{j}(\mathbf{r},t)=\partial_{t}\mathbf{p}(\mathbf{r},t)$.
This implies that conductivity and polarizability tensors in the frequency domain are related in such a way that
\begin{equation}
 \bm{\sigma}^{(0)}_{1}(\omega)=-i\omega\bm{\alpha}^{(0)}_{1}(\omega),
\end{equation}
and
\begin{equation}
 \bm{\sigma}^{(0)}_{2}(\omega_{1},\omega_{2})=-i(\omega_{1}+\omega_{2})\bm{\alpha}^{(0)}_{2}(\omega_{1},\omega_{2}),
\end{equation}
at first and second order, respectively.
Making the pertinent substitutions, we finally write the Dyson-like linear and quadratic optical polarization-density response equations as
\begin{equation}\label{eq:alpha11}
 \bm{\alpha}_{1}(\omega)=\bm{\alpha}^{0}_{1}(\omega)\bm{\varepsilon}^{-1}(\omega),
\end{equation}
with $\bm{\varepsilon}^{-1}(\omega)=\bm{\mathbb{1}}-[4\pi-\bm{A}_{\mathrm{LRC}}(\omega)]\bm{\alpha}_{1}(\omega)$ the inverse of the optical dielectric tensor, and
\begin{equation}\label{eq:alpha2}
 \begin{split}
  \bm{\alpha}_{2}(\omega_{1},\omega_{2})=&\bm{\varepsilon}^{-1,T}(\omega_{1}+\omega_{2})\bm{\alpha}^{0}_{2}(\omega_{1},\omega_{2})\bm{\varepsilon}^{-1}(\omega_{1})\bm{\varepsilon}^{-1}(\omega_{2})\\&+i\bm{\alpha}_{1}^{T}(\omega_{1}+\omega_{2})\bm{B}_{\mathrm{LRC}}(\omega_{1},\omega_{2})\bm{\alpha}_{1}(\omega_{1})\bm{\alpha}_{1}(\omega_{2})
 \end{split},
\end{equation}
respectively, free of any kind of divergence as can be seen.
Hence, these are the equations we use in this work in order to account for many-body effects within the optical response up to second order at the microscopic level.
\subsection{From microscopic to macroscopic optical response}\label{subsec:macromicro}
The ultimate goal of this work is thus to calculate the macroscopic optical susceptibility tensor from first principles and compare it with experiment.
This is done by performing a macroscopic average of the interacting optical polarizability tensor at the microscopic level given by Eqs.~\ref{eq:alpha11} and \ref{eq:alpha2}~\cite{PhysRevB.107.205101}.
In atomic units, this task is done by means of the following expressions for first and second order,
\begin{equation}
 \bm{\chi}_{1}(\omega)=4\pi\bm{\alpha}_{1}(\omega)\bm{\varepsilon}_{\mathrm{M}}(\omega)
\end{equation}
and
\begin{equation}
 \bm{\chi}_{2}(\omega_{1},\omega_{2})=4\pi\bm{\varepsilon}^{T}_{\mathrm{M}}(\omega_{1}+\omega_{2})\bm{\alpha}_{2}(\omega_{1},\omega_{2})\bm{\varepsilon}_{\mathrm{M}}(\omega_{1})\bm{\varepsilon}_{\mathrm{M}}(\omega_{2}),
\end{equation}
respectively, where the macroscopic optical dielectric tensor is given by
\begin{equation}\label{eq:em}
 \bm{\varepsilon}_{\mathrm{M}}(\omega)=\left[\bm{\mathbb{1}}-4\pi\bm{\alpha}_{1}(\omega)\right]^{-1}.
\end{equation}

\section{Trionic peaks in doped semiconductors within TDCDFT}\label{sec:apptrions}
Well-defined trionic absorption peaks have been observed in the optical spectra of doped monolayer semiconductors, as reported in Refs.~\cite{Ross2013,Mak2013}.
Although these features appear in linear optical measurements, their microscopic description requires extending the response theory beyond linear order.
Up to second order, the microscopic polarization-density response is given by [see Eq.~(\ref{eq:peqaEext})]
\begin{equation}\label{eq:p1Eext}
\begin{split}
\mathbf{p}_{1}(\omega)&=\bm{\alpha}_{1}(\omega)\mathbf{E}_{\mathrm{ext}}(\omega)\\
\mathbf{p}_{2}(\omega=\omega_{1}+\omega_{2})&=\bm{\alpha}_{2}(\omega_{1},\omega_{2})\mathbf{E}_{\mathrm{ext}}(\omega{1})\mathbf{E}_{\mathrm{ext}}(\omega{2}),
\end{split}
\end{equation}
where $\mathbf{p}_{n}$ and $\bm{\alpha}_{n}$ denote the $n$th-order polarization-density vector and many-body polarizability tensor, respectively.
Their relation to the corresponding independent-particle quantities is given in Eqs.~(\ref{eq:alpha11}) and (\ref{eq:alpha2}) at first and second order, respectively.

\begin{figure}[t]
 \begin{center}
  \includegraphics[width=0.615\linewidth]{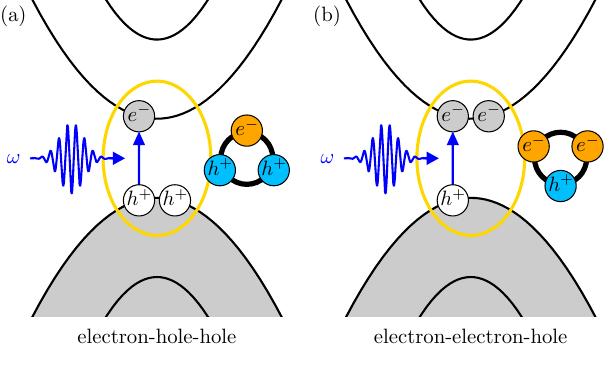}
  \caption{Schematic of three-particle many-body interactions of a gate-voltage induced p-doped (a) and n-doped (b) semiconductor, respectively, when absorbing a photon of frequency $\omega$. Grey and white shaded lines represent valence and conduction bands, respectively.}
  \label{fig:figapp}
 \end{center}
\end{figure}
In optical experiments on doped samples, charge carriers are commonly introduced by applying a gate voltage.
Within TDCDFT, this is described by a static electric field $\mathbf{E}_{\mathrm{g}}(\omega=0)$ acting on the system.
In the presence of this field, the linear electro-optic effect contributes to the optical response.
Although measured as a linear optical response, it is microscopically a quadratic process involving the absorption of a photon with frequency $\omega$, which creates an electron-hole pair, together with the absorption of a zero-frequency photon associated with the static electric field, which either promotes an electron to the conduction band or creates a hole in the valence band, as illustrated in Fig.~\ref{fig:figapp}(a) and (b), respectively.
These processes give rise to electron-hole-hole and electron-electron-hole interactions, respectively, which are accounted for by the second-order xc kernel, as discussed in the introduction.
The corresponding quadratic polarization is therefore
\begin{equation}\label{eq:p2EgEext}
\mathbf{p}_{2}(\omega)=\bm{\alpha}_{2}(\omega)\mathbf{E}_{\mathrm{g}}\mathbf{E}_{\mathrm{ext}}(\omega),
\end{equation}
with $\mathbf{E}_{\mathrm{g}}(\omega=0)\equiv\mathbf{E}_{\mathrm{g}}$.
Up to second order, the microscopic external field is related to the total field through
\begin{equation}\label{eq:Eext2Etotp12}
\mathbf{E}_{\mathrm{ext}}(\omega)=\mathbf{E}_{\mathrm{tot}}(\omega)+4\pi\left[\mathbf{p}_{1}(\omega)+\mathbf{p}_{2}(\omega)\right],
\end{equation}
where atomic units ($4\pi\varepsilon_{0}=1$) are used throughout.
Substituting Eq.~(\ref{eq:Eext2Etotp12}) into Eq.~(\ref{eq:p1Eext}) and Eq.~(\ref{eq:p2EgEext}) yields
\begin{equation}
\begin{split}
\mathbf{p}_{1}(\omega)&=\bm{\alpha}_{1}(\omega)\mathbf{E}_{\mathrm{tot}}(\omega)+4\pi\bm{\alpha}_{1}(\omega)\left[\mathbf{p}_{1}(\omega)+\mathbf{p}_{2}(\omega)\right]\\
\mathbf{p}_{2}(\omega)&=\bm{\alpha}_{2}(\omega)\mathbf{E}_{\mathrm{g}}\mathbf{E}_{\mathrm{tot}}(\omega)+4\pi\bm{\alpha}_{2}(\omega)\left[\mathbf{p}_{1}(\omega)+\mathbf{p}_{2}(\omega)\right].
\end{split}
\end{equation}
Combining these equations, the addition of microscopic polarization-density vectors is expressed as a function of the total electric field as
\begin{equation}
\mathbf{p}_{1}(\omega)+\mathbf{p}_{2}(\omega)=\left\{\bm{\mathbb{1}}-4\pi\left[\bm{\alpha}_{1}(\omega)+\bm{\alpha}_{2}(\omega)\mathbf{E}_{\mathrm{g}}\right]\right\}^{-1}
\left[\bm{\alpha}_{1}(\omega)+\bm{\alpha}_{2}(\omega)\mathbf{E}_{\mathrm{g}}\right]
\mathbf{E}_{\mathrm{tot}}(\omega).
\end{equation}
In the optical limit, the macroscopic average of microscopic quantities is straightforward as long as they are expressed in terms of the total electric field.
Consequently, for doped systems the macroscopic dielectric tensor is defined as
\begin{equation}
\bm{\varepsilon}_{\mathrm{M}}(\omega)=\bm{\mathbb{1}}+\left\{\bm{\mathbb{1}}-4\pi\left[\bm{\alpha}_{1}(\omega)+\bm{\alpha}_{2}(\omega)\mathbf{E}_{\mathrm{g}}\right]\right\}^{-1}
4\pi\left[\bm{\alpha}_{1}(\omega)+\bm{\alpha}_{2}(\omega)\mathbf{E}_{\mathrm{g}}\right],
\end{equation}
which, after some rearrangements, can be simply rewritten as
\begin{equation}\label{eq:em2}
\bm{\varepsilon}_{\mathrm{M}}(\omega)=
\left\{\bm{\mathbb{1}}-4\pi\left[\bm{\alpha}_{1}(\omega)+\bm{\alpha}_{2}(\omega)\mathbf{E}{\mathrm{g}}\right]\right\}^{-1}.
\end{equation}

If the linear electro-optic contribution is neglected, Eq.~(\ref{eq:em2}) reduces to Eq.~(\ref{eq:em}).
Within this approximation, doping modifies $\bm{\alpha}_{1}(\omega)$ through changes in the electronic occupation and band structure, leading to shifts in the exciton binding energies, Pauli blocking of low-energy transitions, and an overall reduction in excitonic absorption. %
However, these effects alone cannot generate an additional absorption resonance.

The appearance of trionic absorption peaks originates from the second-order contribution $\bm{\alpha}_{2}(\omega)\mathbf{E}_{\mathrm{g}}$ in Eq.~(\ref{eq:em2}).
Since $\bm{\alpha}_{2}(\omega)$ contains the quadratic LRC screening tensor $\bm{B}_{\mathrm{LRC}}(\omega)$ [see Eq.~(\ref{eq:alpha2})], it incorporates the underlying three-body correlations responsible for bound-trion formation.
Consequently, within TDCDFT, the additional absorption peaks associated with bound trions can only be described when the second-order xc kernel is included.
\end{appendix}
%
%
%
%
%
\bibliography{main}
%
%
\end{document}